\documentclass{article}

\usepackage[preprint]{neurips_2025}

\usepackage[utf8]{inputenc} 
\usepackage[T1]{fontenc}    
\usepackage{hyperref}       
\usepackage[table]{xcolor} 
\usepackage{booktabs}       
\usepackage{array} 
\usepackage{cellspace} 
\usepackage{url}            
\usepackage{booktabs}       
\usepackage{amsfonts}       
\usepackage{nicefrac}       
\usepackage{microtype}      
\usepackage{xcolor}         
\usepackage{graphicx}

\newcommand{\beq}{\begin{equation}}
\newcommand{\eeq}[1]{\label{#1}\end{equation}}

\newcommand{\mcvar}[1]{\mathop{\mathrm{#1}}}
\newcommand{\windspeed}[1]{\mathop{\mathrm{WS}_{#1}}}

\title{Imposing the Fundamental Dynamical Constraint of Hydrostatic Balance to Improve Global ML Weather Prediction}

\author{%
  Akshay Subramaniam \\
  NVIDIA \\
  Santa Clara, CA \\
  \texttt{asubramaniam@nvidia.com} \\
  \And
  Dale Durran \\
  NVIDIA and University of Washington \\
  Santa Clara, CA  \\
  \texttt{ddurran@nvidia.com} \\
  \And
  David Pruitt \\
  NVIDIA \\
  Santa Clara, CA  \\
  \texttt{dpruitt@nvidia.com} \\
  \And
  Nathaniel Cresswell-Clay \\
  University of Washington \\
  Seattle, WA  \\
  \texttt{nacc@uw.edu} \\
  \And
  William Yik \\
  University of Washington \\
  Seattle, WA  \\
  \texttt{yikwill@uw.edu}
}

\begin{document}

\maketitle

\begin{abstract}

Forecasting weather accurately and efficiently is a critical capability in our ability to adapt to climate change. Data driven approaches to this problem have enjoyed much success recently providing forecasts with accuracy comparable to physics based numerical prediction models but at significantly reduced computational expense. However, these models typically do not incorporate any physics priors. In this work, we demonstrate improved skill of data driven weather prediction approaches by incorporating physical constraints, specifically in the context of the DLWP model \citep{karlbauer_advancing_2024}. Near hydrostatic balance, between the vertical pressure gradient and gravity, is one of the most fundamental and well satisfied constraints on atmospheric motions.  We impose this balance through both hard and soft constraints, and demonstrate that the soft constraint improves the RMSE of many forecast fields, particularly at lead times beyond 7--10 days.  The positive influence of hydrostatic balance is also clearly evident in improving the physicality and strength of a 10-day forecast for hurricane Irma. These results show that adding appropriate physical constraints can improve the skill and fidelity of data driven weather models in a way that does not impose any significant additional memory capacity or scalability challenges.

\end{abstract}

\section{Introduction}

Machine learning (ML) models have demonstrated skill matching or exceeding that of the best traditional numerical prediction models at medium range forecast lead times out to roughly 10 days \cite{bi_accurate_2023,lam_learning_2023,chenFengWuPushingSkillful2023,chenFuXiCascadeMachine2023,lang_AIFS_2024}.  These models generally do not incorporate physical constraints. Stable long autoregressive simulations of the current climate out to 1000 years have also been achieved by ML models without physical constraints \cite{cresswell-clay_Deep_2024}, but most ML models attempting multi-decadal autoregressive simulations have included some basic physical constraints.

As reviewed in \cite{kashinath_Physics_2021}, physics informed ML models may include physical constraints in the model architecture itself (hard constraints), through training in the loss function (soft constraints), or by tying the ML components of the model architecture to a dynamical core that numerically integrates the governing equations at coarse resolution. This last approach includes numerous attempts to create ML modules to replace traditional parameterizations of processes like subgrid-scale deep convection \cite{yu2023climsim, hu2024stable, kochkov_Neural_2023}.  Of particular note is the Neural GCM, in which the dynamical core is differentiable and the fully coupled dynamics and ML modules can be trained together by backpropagation \cite{kochkov_Neural_2023}.  
Among the pure ML models for simulation on climate time scales the ACE2 model \cite{watt-meyer_ACE2_2024} enforces global conservation of the dry air mass and total water, as well as the column integrated moisture.

We are not aware of a pure ML weather or climate model that enforces local constraints arising from the momentum equations.  
Hydrostatic balance, between the vertical pressure gradient and gravitational acceleration $g$, is the most fundamental and well-satisfied  balance in atmospheric dynamics.  Even the small fraction of the total thermodynamic pressure responsible for horizontal pressure gradients and motions is very nearly in hydrostatic balance for circulations with horizontal length scales greater than 20 km \cite{vallis_2017}.

In the following we present a physics constrained model that preserves approximate hydrostatic balance and show that this constraint improves several measures of model performance, particularly beginning at forecast lead times of 7--10 days.

\section{Methods}
\subsection{Hydrostatic Balance}
\label{sec:hydrostatic-balance}

Assuming hydrostatic balance, the vertical momentum equation can be approximated as 
\beq \frac{1}{\rho}\frac{dp}{dz}=-g, \eeq{eqn:hydrostatic}
where $p$ is pressure, $\rho$ density, $g$ gravitational acceleration, and $z$ is the vertical coordinate.
Using the equation of state and the virtual temperature
\[ T_v=T(1+ 0.6078\,q), \]
which accounts for the influence of water vapor on density through the specific humidity $q$, (\ref{eqn:hydrostatic}) may be expressed as
\beq RT_v\frac{d(\ln p)}{dz} = -g, \eeq{eqn:p_hydrostatic}
where $R$ the gas constant for dry air. Switching to pressure as the vertical coordinate, defining the
geopotential height $Z(p)$ as the height above mean sea level where the atmospheric pressure is $p$, and integrating between a lower level denoted by ``1'' and an upper level ``2'', (\ref{eqn:p_hydrostatic}) becomes
\beq \int_{p_1}^{p_2}T_v \, d(\ln p) = -\frac{g}{R} \left( Z_2-Z_1\right ). \eeq{eqn:hypsometric}
Using a trapezoidal approximation to the integral, the average $T_v$ across the layer satisfies
\beq \frac{T_v(p_1) + T_v(p_2)}{2}= \frac{g}{R\ln (p_1/p_2)}\left ( Z_2-Z_1 \right ) .
\eeq{eqn:FD_hypsometric}
Equation~\ref{eqn:FD_hypsometric} can also be derived in differential form directly by discretizing the derivative in (\ref{eqn:p_hydrostatic}) using a $2^{\mathrm{nd}}$ order finite-difference approximation at the mid-point of levels ``1'' and ``2'' and averaging $T_v$ between those levels.

\subsection{Imposing a hard constraint}

Noting that $p_1$ and $p_2$ are specified values of the vertical coordinate, (\ref{eqn:FD_hypsometric}) is a linear relation between geopotential height and virtual temperature, and as suggested in \cite{beucler_enforcing_2021}, it can be incorporated as a hard constraint in the model architecture and included in backpropagation over each training cycle.  We tried this approach by training a model to prognostically determine all forecast variables at the next time step, except the virtual temperature at all heights above 850 hPa. The upper-level values of $T_v$ were computed by solving
\beq T_v(p_2)= \frac{2g}{R\ln(p_1/p_2)}\left (Z_2-Z_1\right ) -T_v(p_1), \eeq{eqn:constraint}
starting with $(p_2,p_1)=(700,850)$ hPa and iteratively progressing upward through all vertical levels in each grid cell. We start the iteration at 850 hPa instead of 1000 hPa to reduce the number of special cases in which $Z_1$ lies below the topography.

This approach failed because of numerical errors, including the accumulation of error as the iteration proceeds upwards, coarse vertical grid spacing, and the low-order accuracy of trapezoidal quadrature. The balance between the full vertical pressure gradient and gravity is so exacting that it cannot be readily included in discrete models of atmospheric motion, so approximate equations are developed that avoid it's explicit computation. For large-scale motions, the approximate governing equations assume hydrostatic balance, even for the small horizontally varying fraction of the full pressure field that drives horizontal motions.  On smaller scales with horizontal grid spacing less than about 10 km, the dynamically relevant pressures and densities are defined as perturbations about a specified hydrostatically balanced vertical profile, but moderate inaccuracies in that balance do not feed back on the numerical solution.

\subsection{Imposing a weak constraint}
\label{sec:weak-constraint}

\begin{figure}
  \includegraphics[width=0.4\textwidth, keepaspectratio]{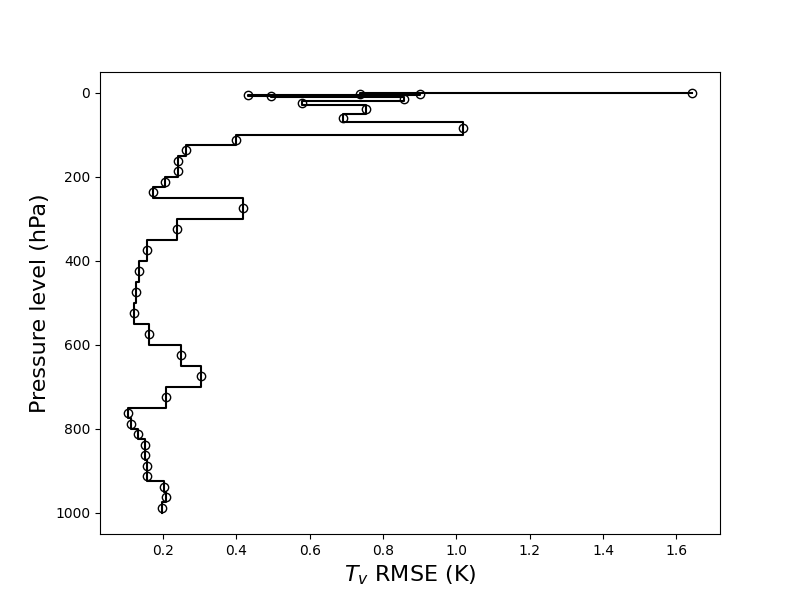}
  \includegraphics[width=0.4\textwidth, keepaspectratio]{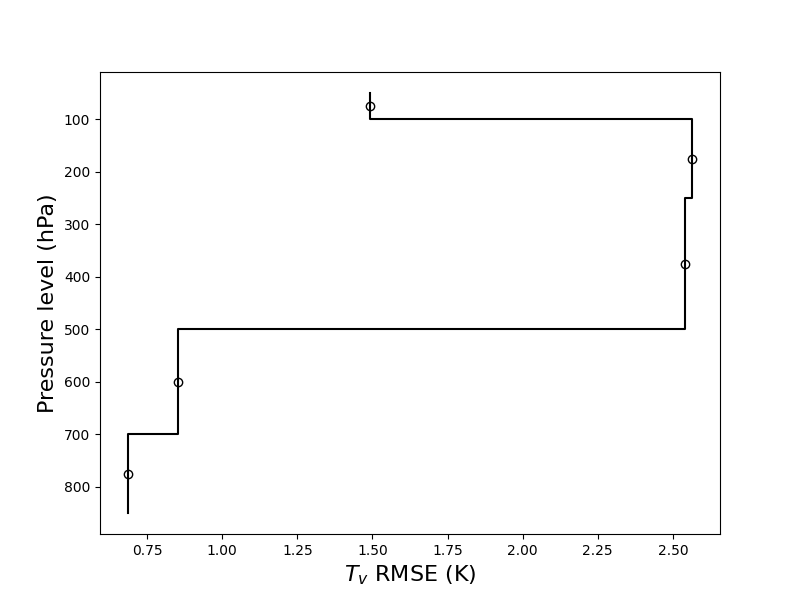}
  \centering
  \caption{Virtual temperature deviations from hydrostatic balance, plotted with pressure as the height coordinate, computed for ERA5 data with all 37 vertical levels (left) and the six vertical levels above 1000 hPa in our model (right). Steps in the plots indicate the pressure-level slabs across which virtual temperature errors are computed using (\ref{eqn:FD_hypsometric}).}
  \label{fig1}
\end{figure}

ECMWF's Integrated Forecast System (IFS), used to create the ERA5 reanalysis \cite{hersbach2020era5} is a hydrostatic model, and the dynamical constraints it imposes on the reanalysis do not require strict hydrostatic balance. Figure~\ref{fig1} shows the RMSD between the right side of (\ref{eqn:constraint}) and $T_v$ averaged across each of 37 vertical levels (left side of (\ref{eqn:constraint})) in the full-vertial-resolution ERA5 data.  The values in 
Fig.~\ref{fig1} are averaged over the globe and ten time stamps chosen randomly from 2024.  The errors at low levels in the full-resolution ERA5 data are around 0.2 K, which would be very significant in a model that attempted to compute vertical accelerations using all terms in the momentum equations, but is not large compared to observational error.  Owing to our coarse vertical resolution, the RMSDs in our model are significantly larger, particularly above 500 hPa. These differences could be reduced by adding vertical resolution, but here we pursue a different strategy by keeping our model relatively parsimonious, with just six vertical levels, to demonstrate how enforcing even approximate hydrostatic balance can improve model performance.

As suggested by the RMSD magnitudes in Fig.~\ref{fig1}b, it is not possible to both impose a strict hydrostatic constraint and minimize the RMSE between the ERA5 temperatures, humidities and geopotential heights and the corresponding fields predicted by DLESyM.  Therefore, we impose hydrostatic balance through a soft constraint that penalizes the training loss where the imbalance is most pronounced.  

During training, we apply the error-tolerant loss function 
\beq
f(r_k) = \frac{\left(r_k/\alpha_k\right)^2}{1 + e^{1-\left(r_k/\alpha_k\right)^2}}
\eeq{eqn:loss}
to the deviation from hydrostatic balance between levels $k$ and $k-1$ computed as the residual
\beq
r_k=\frac{T_v(p_k) + T_v(p_{k-1})}{2}- \frac{g}{R\ln (p_{k-1}/p_k)}\left ( Z_k-Z_{k-1} \right )  .
\eeq{eqn:residual}
The total training loss is scaled MSE plus a Lagrange multiplier times the loss at each level.  For $r_k\le\alpha_k$, $f(r_k)\approx 0$ and the hydrostatic constraint is small compared to the MSE training loss in the prognostic variables.  In contrast, the hydrostatic constraint becomes significant for $r_k>\alpha_k$ as $f(r_k)$ approaches $(r_k/\alpha_k)^2$. Figure~\ref{fig:Tv-error-loss} shows this error tolerant loss function compared to an MSE loss.

\section{Model and Training}

\begin{figure}
  \includegraphics[width=0.9\textwidth, keepaspectratio]{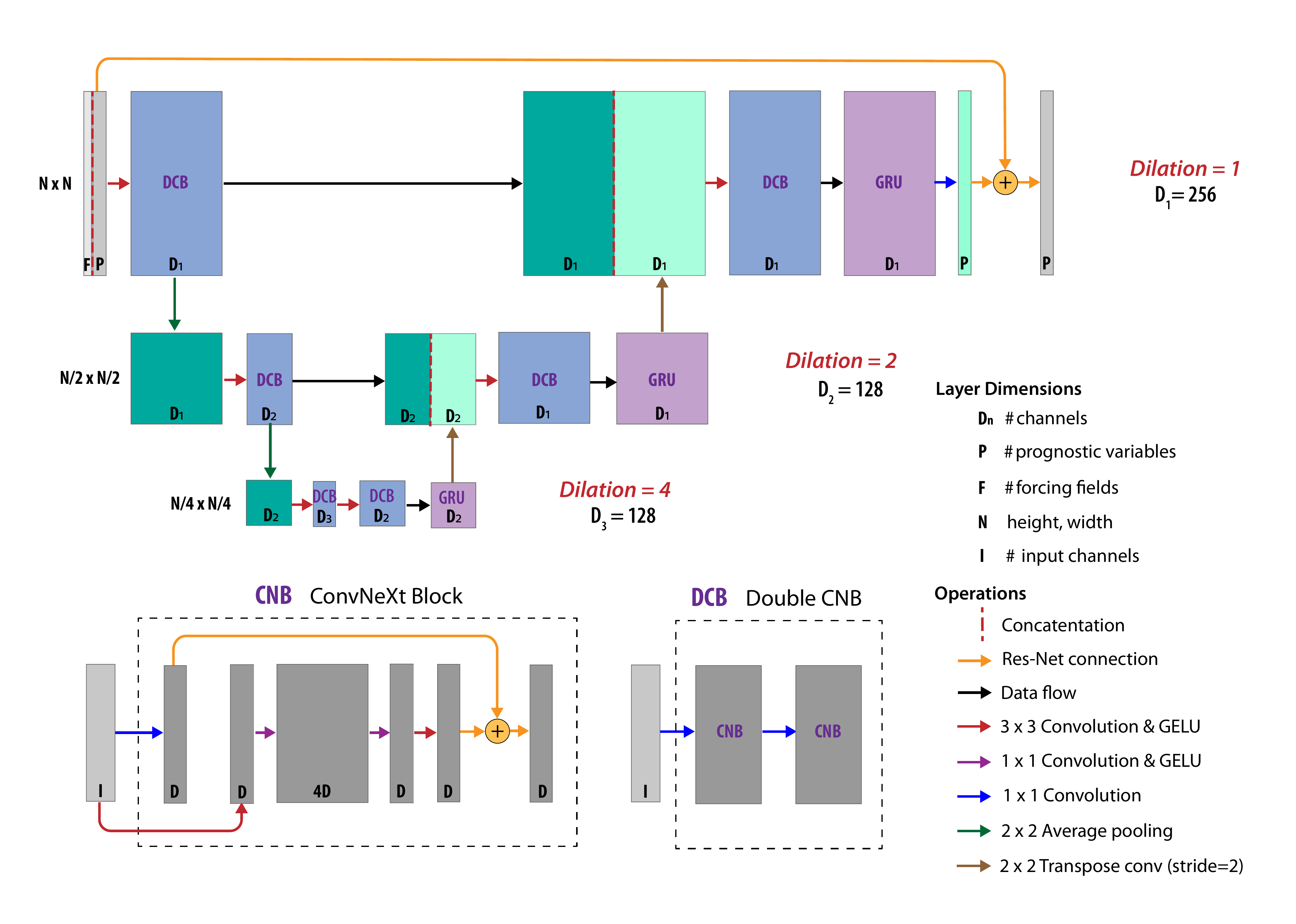}
  \centering
  \caption{Schematic representation of the model architecture as a sequence of operations on layers (see legend). U-net levels are labeled by their channel depth, with $D_1=256$ and $D_2=D_3=128$ being associated with the first convolutions in each level. Each Double ConvNeXt block (blue) is replaced by the layers and operations shown in the inset labeled DCB, with generic embedding depths $D$ and $I$ determined by the channel depth of the input and the labeled value of $D_n$. The purple blocks labeled GRU denote convolutional Gated Recurrent Unit layers, which are implemented with $1\times 1$ spatial convolutions. Other layers evaluated by the encoder are shown as dark green, while those evaluated by the decoder are shown as light green. }
  \label{fig:U-net}
\end{figure}

The model architecture is diagrammed in Fig.~\ref{fig:U-net}, with details of the convolutional GRU omitted for simplicity. The overall structure is similar to that in \cite{karlbauer_advancing_2024} with data carried on a Hierachical Equal Area isoLatitude Pixelization (HEALPix) \cite{gorski_healpix_2005} having $64\times64$ points on each of its 12 faces.  In contrast to \cite{karlbauer_advancing_2024}, the number of prognostic variables is increased from 9 to 26 (as listed in Table~\ref{prog_variables}), the ConvNeXt blocks are reconfigured as in \cite{cresswell-clay_Deep_2024}, and each block is doubled. The model is implemented in the PhysicsNeMo package \cite{PhysicsNeMo_Contributors_NVIDIA_PhysicsNeMo_An_2023}.

\subsection{Baseline model}
\label{sec:baseline-model}
We train a baseline model trained on just the ERA5 dataset with no hydrostatic constraint as a control. The model is trained on three hourly ERA5 data \cite{hersbach2020era5} from 1980 to 2016 with 2016 to 2019 forming the validation set. We use geopotentials and temperatures at 6 vertical levels from 850 hPa to 50 hPa to be able to evaluate the enforcement of hydrosatic balance. Specific humidity is only used up to 500 hPa heights and assumed to be identically zero at heights above that. In addition to prognostic variables, we also use two prescribed fields: a land-sea fraction and the surface geopotential. A full list of variables used in the model is listed in Table~\ref{prog_variables}.

The model has 8.5 M parameters and is trained for 400 epochs over 16 hours on 64 NVIDIA H100 GPUs with 80 GB HBM memory capacity each. A batch size of 8 per GPU is chosen resulting in a global batch size of 512. Following \cite{karlbauer_advancing_2024}, we use the Adam optimizer \cite{kingma2014adam} and a cosine annealing learning rate schedule with the initial learning rate set to $6 \times 10^{-4}$, the minimum learning rate set to $4 \times 10^{-5}$ and an annealing period of 650 epochs. We also use gradient clipping to stabilize training by clipping gradients to have a maximum L2-norm of $0.25$. The model is trained using a weighted MSE loss for each prognostic variable with Lagrange multipliers chosen to ensure the contribution to overall loss between the smallest and largest individual variable is less than a factor of five.

\begin{table}[htb]
    \centering
        \caption{Prognostic variables carried at each model level.  Here $T$ is temperature, WS wind speed, $q$ specific humidity, $Z$ geopotential height, and TCWV total column water vapor. Subscripts indicate the vertical level.}
    \begin{tabular}{SlSl}
        \hline
        \textup{Vertical level} & Variables \\
        \hline
        \rowcolor{gray!10}
        surface & $T_{2m}$, \, $\windspeed{10m}$, \, $q_{2m}$  \\
        vertically integrated & $\mcvar{TCWV}$  \\
        \rowcolor{gray!10}
        1000 hPa & $Z_{1000}$     \\
        850 hPa & $Z_{850}$, \, $T_{850}$, \, $\windspeed{850}$, \, $q_{850}$  \\
        \rowcolor{gray!10}
        700 hPa & $Z_{700}$, \, $T_{700}$, \, $\windspeed{700}$, \, $q_{700}$  \\
        500 hPa & $Z_{500}$, \, $T_{500}$, \, $\windspeed{500}$, \, $q_{500}$  \\
        \rowcolor{gray!10}
        250 hPa & $Z_{250}$, \, $T_{250}$, \, $\windspeed{250}$  \\
        100 hPa& $Z_{100}$, \, $T_{100}$, \, $\windspeed{100}$  \\
        \rowcolor{gray!10}
        50 hPa  & $Z_{50}$, \, $T_{50}$, \, $\windspeed{50}$ \\
        \hline
        \\
    \end{tabular}
    \label{prog_variables}
\end{table}

\subsection{Hydrostatic finetuning}
\label{sec:hydrostatic-finetuning}
We incorporate the weak form of hydrostatic constraints as outlined in Section~\ref{sec:weak-constraint} by finetuning the baseline model with these constraints added to the loss function with the error-tolerant loss (\ref{eqn:loss}) and Lagrange multipliers to weight the additional hydrostatic loss terms. Since the ERA5 training data have inherent hydrostatic balance errors (Fig.~\ref{fig1}), the error-tolerant loss is used to relax the competition between satisfying MSE and hydrostatic losses. We choose the values of $\alpha_k$ per vertical slab to penalize the largest hydrostatic errors while minimizing the penalty for a percentile $p$ of the training data with the smallest hydrostatic imbalances. If $Q_k(p)$ is the imbalance distribution of $T_v$ in the training data at slab $k$, then we choose $\alpha_k$ such that
\beq \left. \frac{\mathrm{d}}{\mathrm{d} r_k}f(r_k) \right|_{r_k=Q_k(p)} = \frac{\mathrm{d}}{\mathrm{d} r_k} \left. \left( \frac{r_k}{\alpha_k} \right)^2 \right|_{r_k=Q_k(p)} . \eeq{eqn:alpha-condition}
Below the point defined by the above condition, the gradients of the error tolerant loss are smaller than an MSE loss while above that point, the gradients are larger than an MSE loss. 
All errors larger than $Q_k(p)$ would be penalized more significantly while the penalty for errors smaller than $Q_k(p)$ would be diminished compared to an MSE loss. In this way, the choice of $p$ determines how strictly hydrostatic balance is enforced in the model. Solving \ref{eqn:alpha-condition} for $\alpha_k$, we get
\beq \alpha_k = \frac{Q_k(p)}{\sqrt{W_0(1) + 1}} \eeq{eqn:alpha-solution}
where $W_0()$ is the principal branch of the Lambert W function \cite{corless1996lambert}.

\begin{figure}
\includegraphics[width=0.4\textwidth, keepaspectratio]{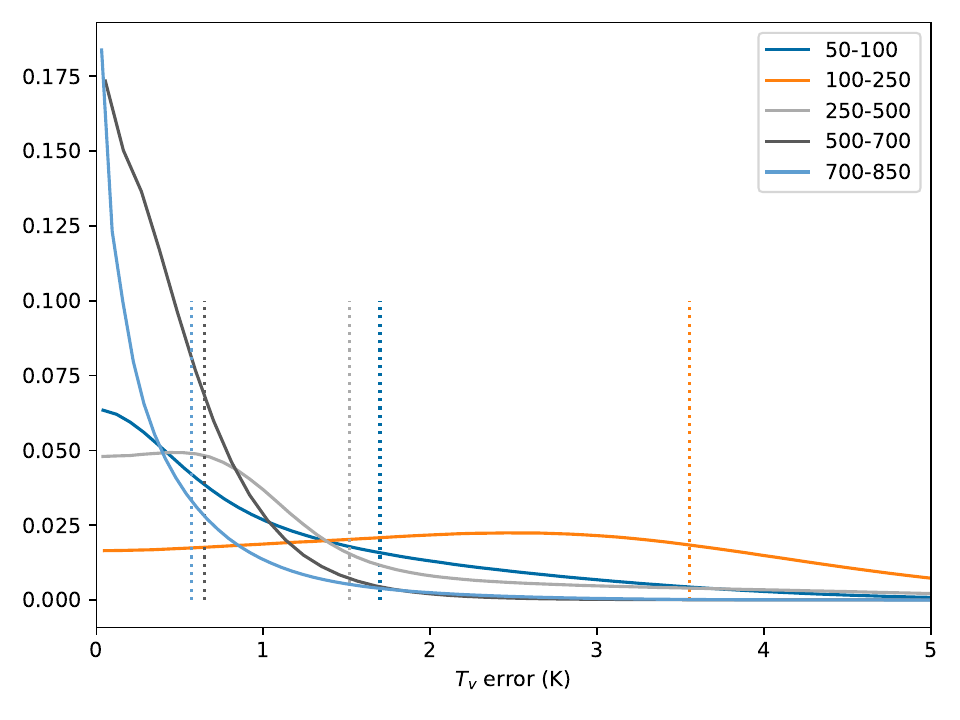}
\includegraphics[width=0.44\textwidth, keepaspectratio]{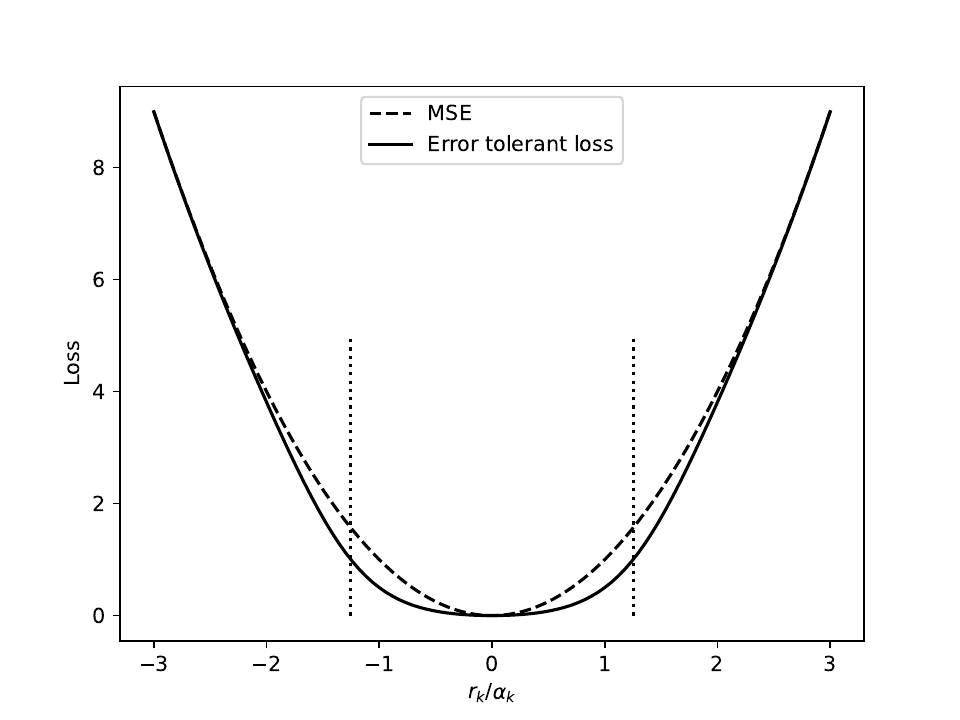}
  \centering
  \caption{Left: Distribution of ERA5 virtual temperature errors compared to hydrostatic balance for the five vertical slabs used in our model in solid lines. Dotted lines indicate the error level of the 75th percentile of the distribution $Q_k(0.75)$. Right: The error tolerant loss function defined in (\ref{eqn:loss}) in solid, dashed line shows a regular MSE loss function and the vertical dotted lines indicate the point at which the derivatives of both functions are equal (\ref{eqn:alpha-condition}). The vertical dotted lines on the two panels are matched to set $\alpha_k$ for each vertical slab (\ref{eqn:alpha-solution}).}
  \label{fig:Tv-error-loss}
\end{figure}

Figure~\ref{fig:Tv-error-loss} shows a plot of the distribution of the hydrostatic imbalance $| r_k |$ evaluated on 1000 random times of the ERA5 dataset for the five vertical slabs used in our model along with the $Q_k(0.75)$ error values. We fine tune the baseline model using hydrostatic losses for three values of $p$: $0.5$, $0.75$ and $0.95$. Setting $p=0.5$ means that the upper half of the error distribution is penalized more strictly while the lower half has diminished penalty during training. On the other hand, setting $p=0.95$ strongly penalizes only the largest 5\% of hydrostatic errors in the training data. Table~\ref{tab:alphas} shows the $\alpha_k$ values for these three different $p$ values and all the vertical slabs used to enforce hydrostasy. From here on, we will refer to the models finetuned with these three $p$ values as "q95", "q75" and "q50" respectively.

\begin{table}[htb]
    \centering
    \caption{Table outlining the different $\alpha_k$ values used for the three finetuning experiments with $p$ 0.5, 0.75 and 0.95. The last column also shows the Lagrange multiplier used for the hydrostatic loss for each vertical slab.}
    \begin{tabular}{SlSlSlSlSl}
        \hline
        \textup{Vertical Slab}        & $\left. \alpha_k \right| Q(0.5)$    & $\left. \alpha_k \right| Q(0.75)$    & $\left. \alpha_k \right| Q(0.95)$ & Loss Weight \\
        \hline
        \rowcolor{gray!10}
        850-700 hPa & $0.211$ & $0.475$ & $1.213$ & $10^{-3}$ \\
        700-500 hPa & $0.255$ & $0.510$ & $0.957$ & $10^{-3}$ \\
        \rowcolor{gray!10}
        500-250 hPa & $0.640$ & $1.216$ & $3.391$ & $10^{-4}$ \\
        250-100 hPa & $1.895$ & $2.757$ & $4.135$ & $10^{-4}$ \\
        \rowcolor{gray!10}
        100-50 hPa  & $0.676$ & $1.420$ & $2.773$ & $10^{-4}$ \\
        \hline
        \\
    \end{tabular}
    \label{tab:alphas}
\end{table}

\begin{figure}
\includegraphics[width=\textwidth, keepaspectratio]{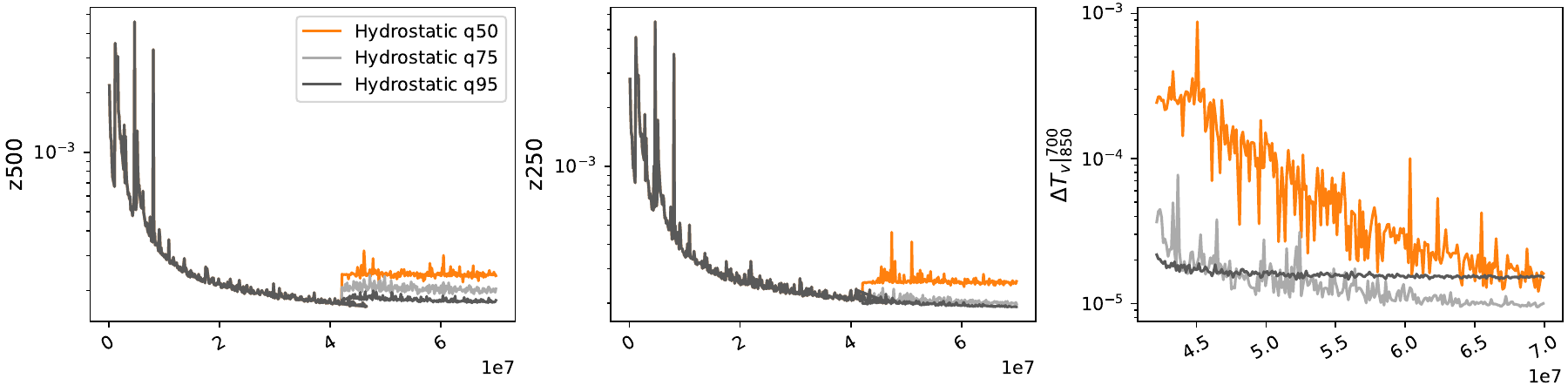}
  \centering
  \caption{Validation losses for $Z_{500}$ and $Z_{250}$ MSE, and for $T_v$ hydrostatic imbalance between 700 and 850 hPa as a function of the training iteration. Finetuning with the hydrostatic losses starts at $\approx 4 \times 10^{7}$ iterations, with the different colors indicating different error quantiles used to set $\alpha_k$.}
  \label{val_losses}
\end{figure}

Figure~\ref{val_losses} shows the validation losses for $Z_{500}$, $Z_{250}$ and hydrostatic errors between 700 and 850 hPa as a function of the training iteration. We see that for $Z_{500}$ and $Z_{250}$, finetuning with the hydrostatic losses increases the MSE validation losses as the hydrostatic constraint penalty is made stronger (q95 to q75 to q50). However, as we will demonstrate in Section~\ref{sec:error-statistics}, these validation losses computed at training time over a single time step are not directly correlated to longer time autoregressive skill indicating that the absolute magnitude of a single-time-step error is less important than the magnitude of error orthogonal to the hydrostatic manifold. 

The hydrostatic-imbalance validation loss shows better convergence with stricter penalization of hydrostasy errors, as expected. The validation loss for hydrostatic imbalance in the 500-700 hPa layer, trains similarly to that for 700-850 Pa shown in Fig.~\ref{val_losses}.  However, the imbalance loss is not reduced during training over the three higher layers, where poor numerical resolution gives larger imbalances when evaluating $r_k$ directly from the ERA5 data (Fig.~\ref{fig1}).  The hydrostatic imbalance at upper levels can be reduced in magnitude and its loss improved through training by increasing the vertical resolution.  Nevertheless, for our purpose it is instructive to see how the models adjust toward hydrostatic balance during autoregressive rollouts at the current coarse resolution.

\section{Error Statistics}
\label{sec:error-statistics}

\begin{figure}
\includegraphics[width=0.32\textwidth, keepaspectratio]{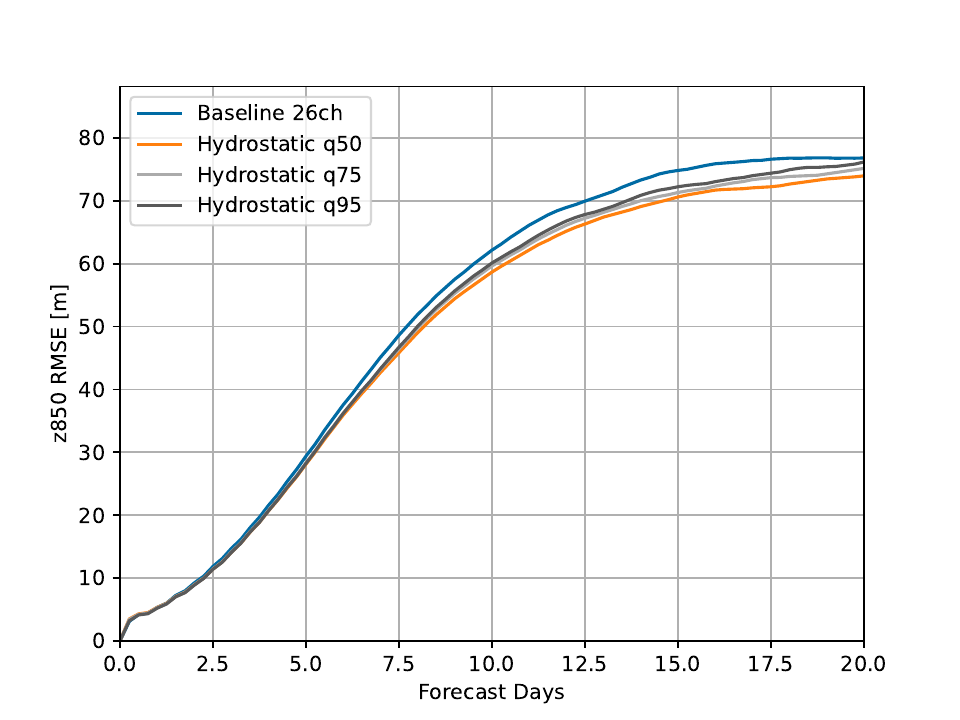}
\includegraphics[width=0.32\textwidth, keepaspectratio]{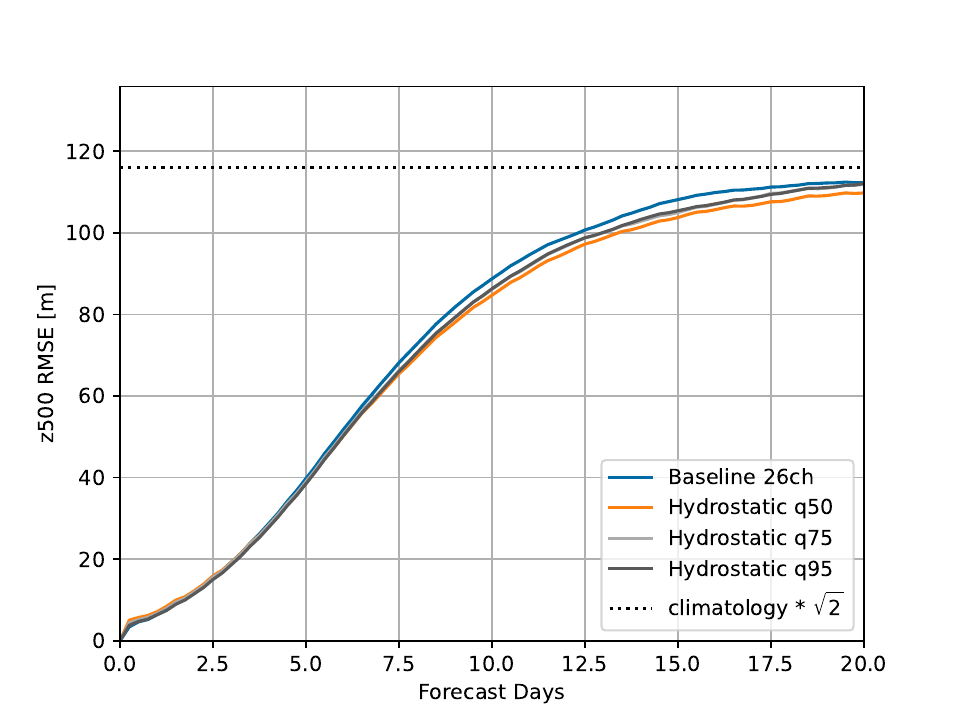}
\includegraphics[width=0.32\textwidth, keepaspectratio]{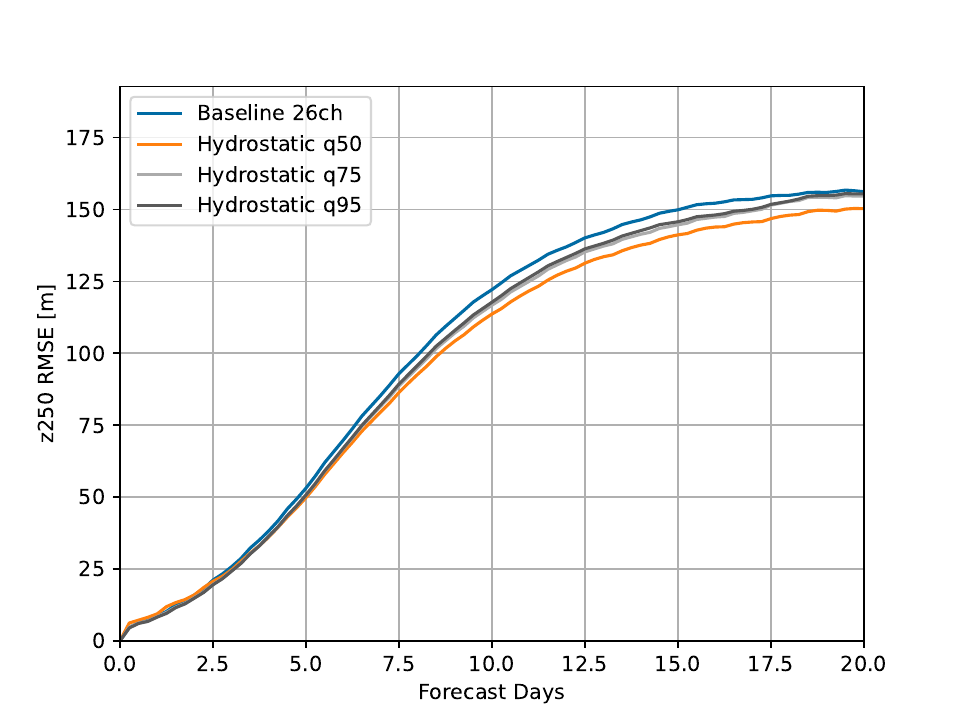}
\includegraphics[width=0.32\textwidth, keepaspectratio]{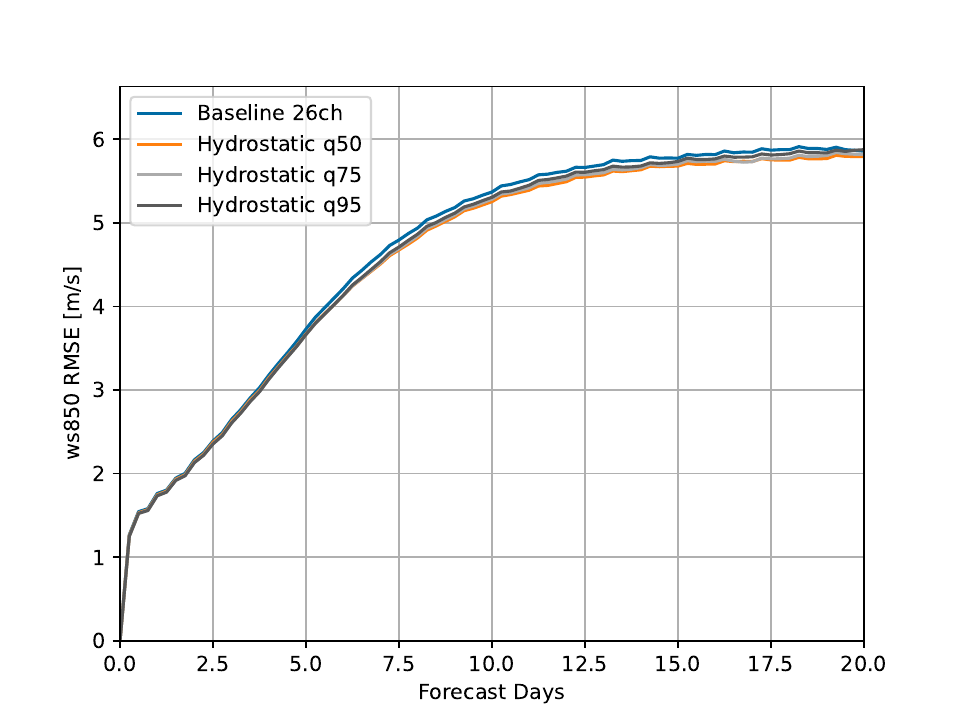}
\includegraphics[width=0.32\textwidth, keepaspectratio]{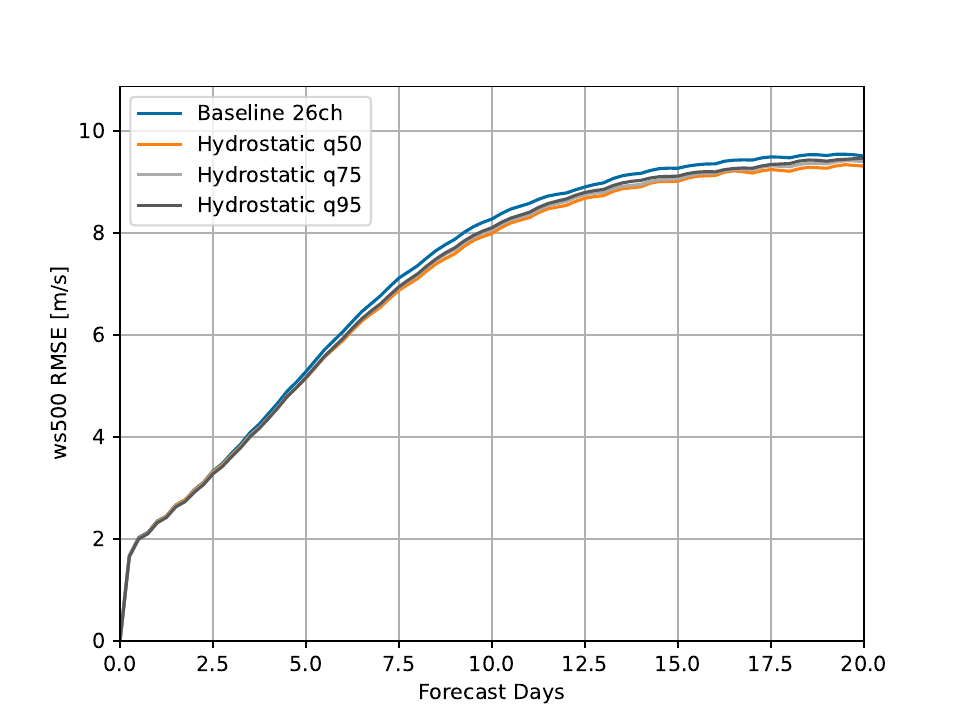}
\includegraphics[width=0.32\textwidth, keepaspectratio]{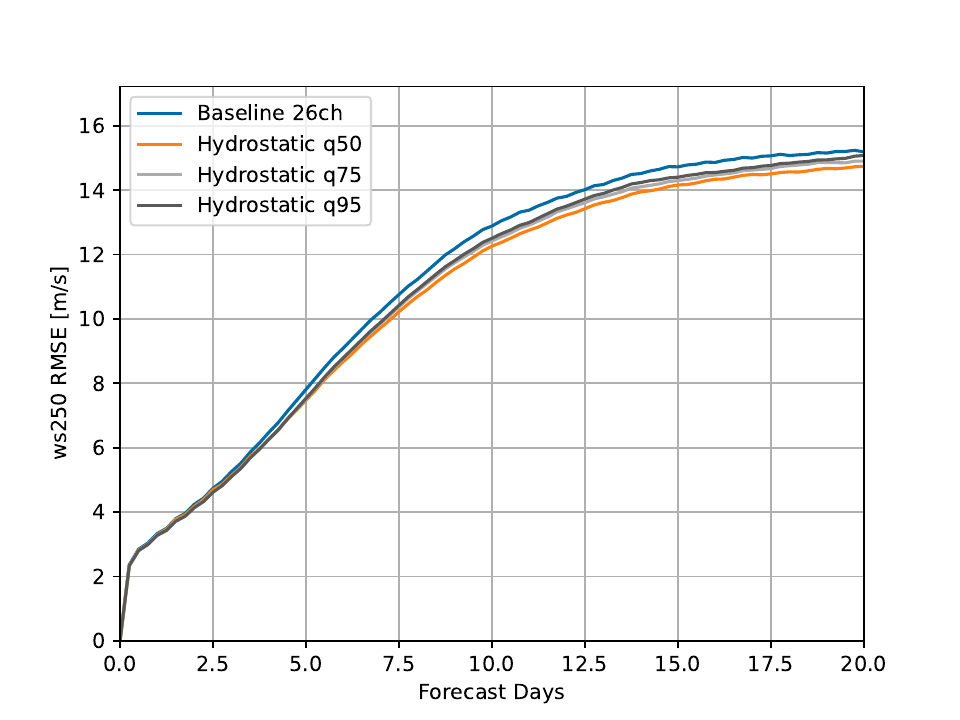}
  \centering
  \caption{Comparison of global RMSE curves between the baseline non-hydrostatic model and models with the hydrostatic constraint imposed above the 50, 75 and 95 percentiles. Variables are $Z_{850}$, $Z_{500}$, $Z_{250}$, $\windspeed{850}$, $\windspeed{500}$, and $\windspeed{250}$ (left to right, top to bottom).}
 \label{rmse-curves}
\end{figure}

The tests were carried out by conducting twice-weekly forecasts for the years 2017 and 2018.  RMSE for geopotential height and wind speeds at 850, 500 and 250 hPa is presented in Fig.~\ref{rmse-curves}.  In all cases, the baseline model performs the worst, 
and enforcing the hydrostatic constraint produces substantial improvement in the geopotential height fields beginning around 7 days. This improvement is generally monotonic as the strength of the constraint increases (smaller values of $\alpha_k$, or equivalently, imposing the penalty for imbalance across a larger percentile of the forecast distribution).  The best results are obtained using q50 ($p = 0.5$). 

At 850 hPa, the bottom of the layer across which the hydrostatic-balance trains to the lowest validation loss, the 60-m error threshold in $Z_{850}$ RMSE is crossed about one day later in the q50 forecasts than in the baseline.  Similar but less pronounced improvements also occur in WS$_{850}$, whereas there is almost no change in RMSE for $T_{850}$ and $q_{850}$ (not shown).  The situation is similar, though with slightly less improvement, at 500 hPa. This improvement is suprising because, in contrast to $Z_{850}$, the validation loss in $Z_{500}$ RMSE is clearly degraded by our fine tuning, particularly for the q50 case (Fig.~\ref{val_losses}), which nevertheless continues to give the best performance in actual tests.

Even more surprising is the substantial improvement of the q50 model relative to the baseline at 250 hPa. At this level, fine-tuning again degrades the RMSE in $Z_{250}$ (Fig.~\ref{val_losses}), and the validation loss for the hydrostatic constraint in the layers above and below 250 hPa does not decrease significantly during training.  Yet WS$_{250}$ (Fig.~\ref{rmse-curves}) and even $T_{250}$ (not shown) in are clearly superior in the q50 model.  Enforcing hydrostatic balance also improves the ACC scores in the test set, with the q50 model again giving the best results (not shown), but  
the improvement is modest.

\begin{figure}
  \includegraphics[width=0.32\textwidth, keepaspectratio]{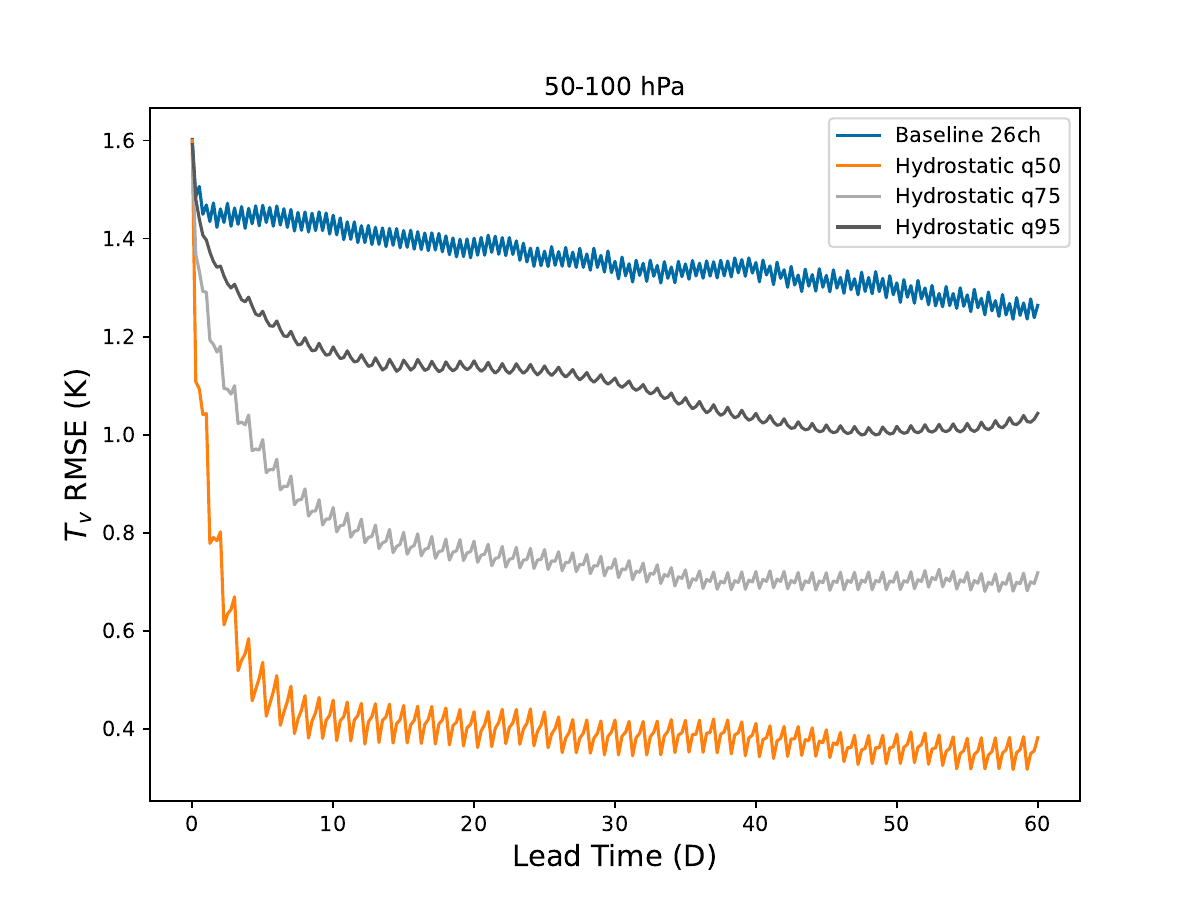}
  \includegraphics[width=0.32\textwidth, keepaspectratio]{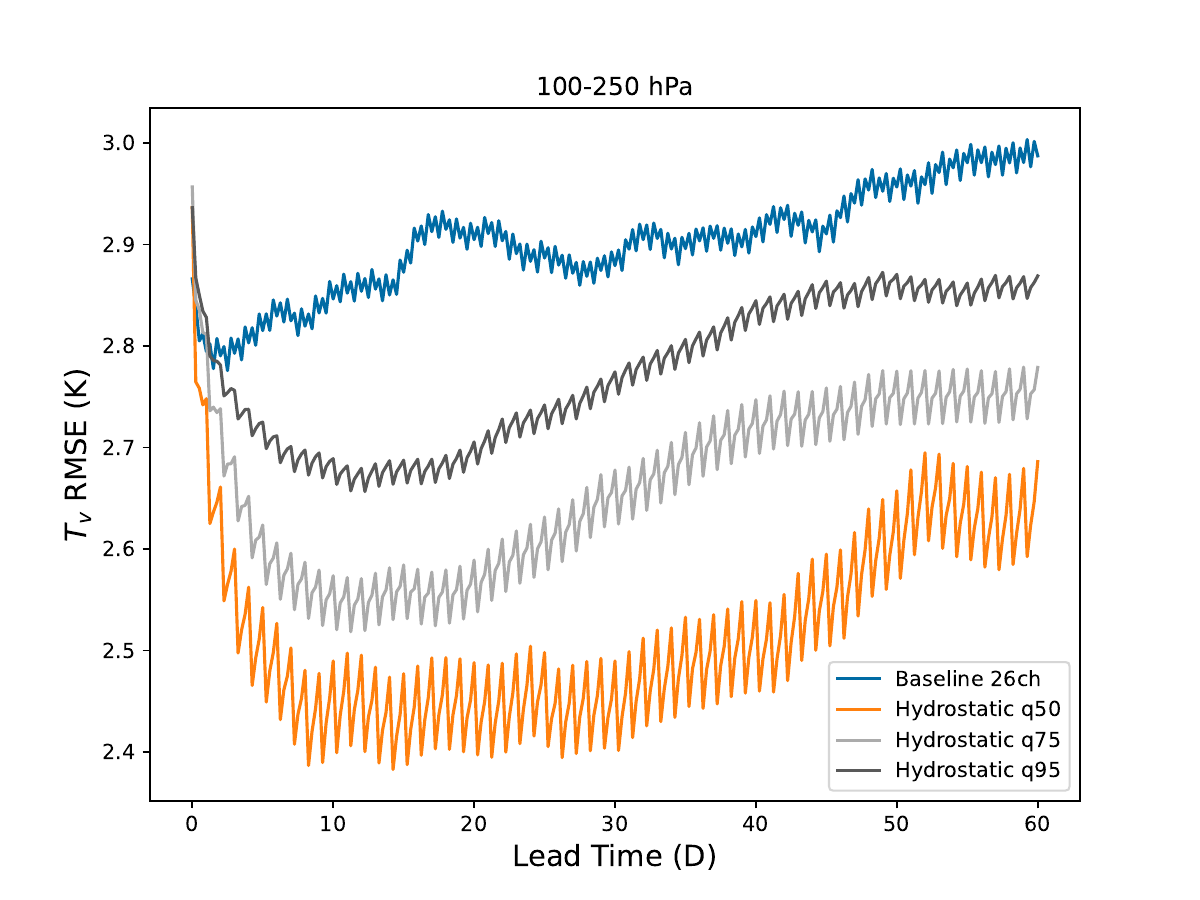}
  \includegraphics[width=0.32\textwidth, keepaspectratio]{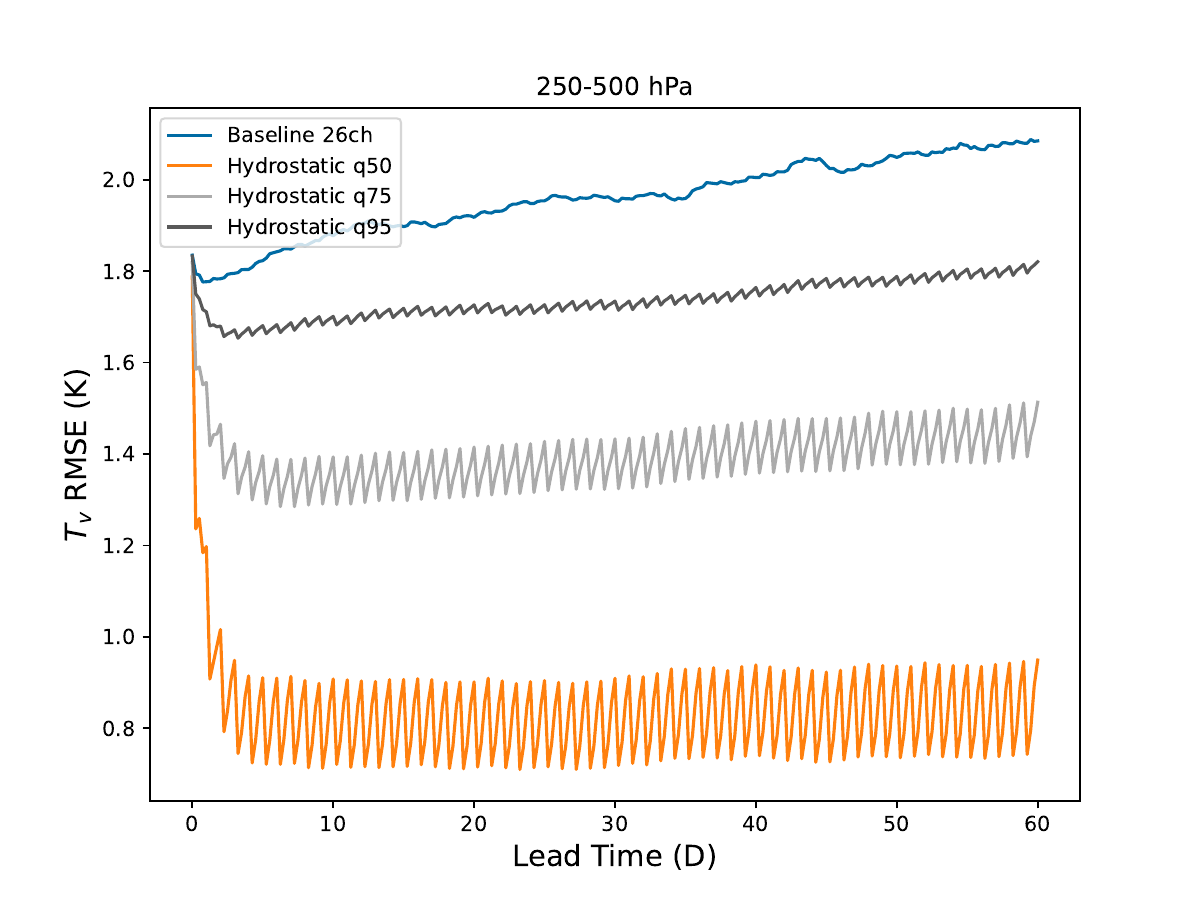}
  \includegraphics[width=0.32\textwidth, keepaspectratio]{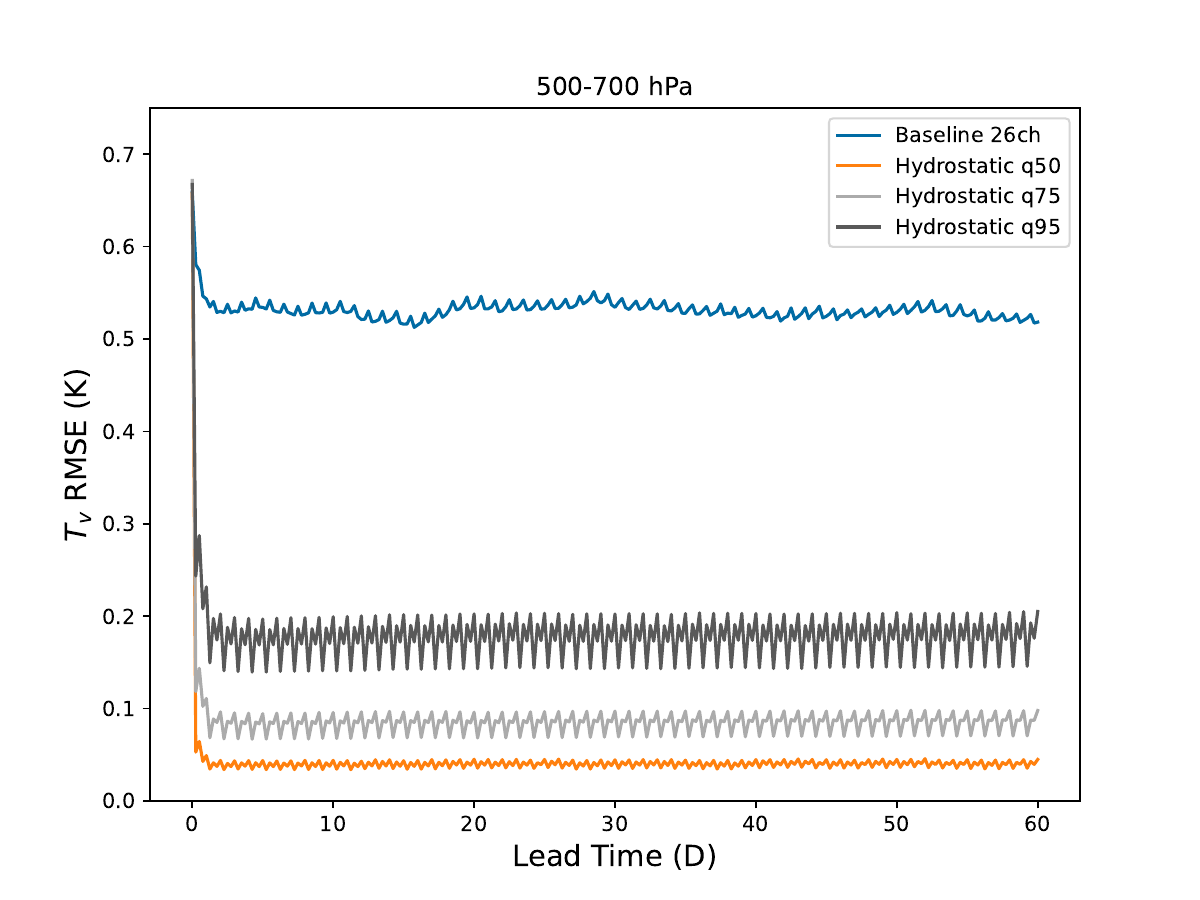}
  \includegraphics[width=0.32\textwidth, keepaspectratio]{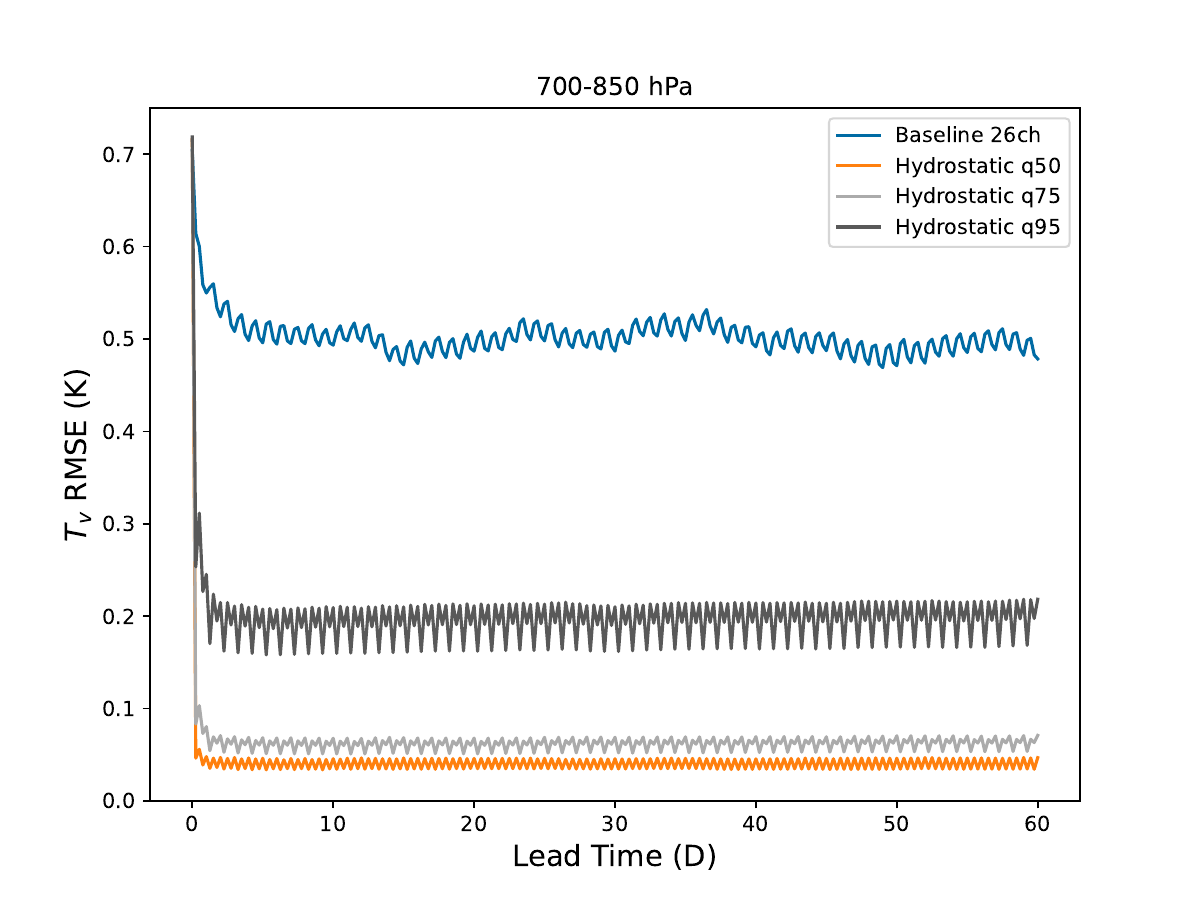}
  \centering
  \caption{Magnitude of the hydrostatic imbalance $|r_k|$ across the model layers as a function of time for the baseline, q50, q75 and q95 models. Note the different scales on the ordinate in each panel.}
  \label{imbalance_by_layer}
\end{figure}

The test-set averaged evolution of the hydrostatic imbalance in each model layer is plotted as a function of time for a 60-day autoregressive rollout in Fig.~\ref{imbalance_by_layer}.  The degree of imbalance is significantly reduced over the first few days in all layers. In the lowest two layers, where it was possible to train the validation loss to minimize hydrostatic imbalance, $|r_k|$ rapidly drops from roughly 0.7 K to less than 0.05 K in the q50 forecasts, a value below that computed using the full-vertical-resolution ERA5 data.  Even in the much more lenient q95 forecasts, it drops to about 0.2 K, with the q75 model performing in between q50 and q95.

Despite being unable to train the model to significantly reduce the validation loss due hydrostatic imbalance in the upper three layers, the imbalance in the 50--100-hPa  and 250--500-hPa layers is greatly reduced over the first few forecast days.  Even the balance for the 100--250-hPa  layer improves significantly, though in contrast to the other four layers, imbalance increases again later in the forecast.

\section{Case study}

As apparent in Fig.~\ref{rmse-curves}, on average, differences between the baseline and hydrostatically constrained models become more pronounced after about 10 days.  Tropical cyclones (TC) can be both intense, and, compared to other synoptic phenomena, long lived. As such, TCs present an opportunity to compare similar 10-day  forecasts from different model configurations without the forecast differences being dominated by hard-to-predict changes in the overall flow.

\begin{figure}
\includegraphics[width=0.9\textwidth, keepaspectratio]{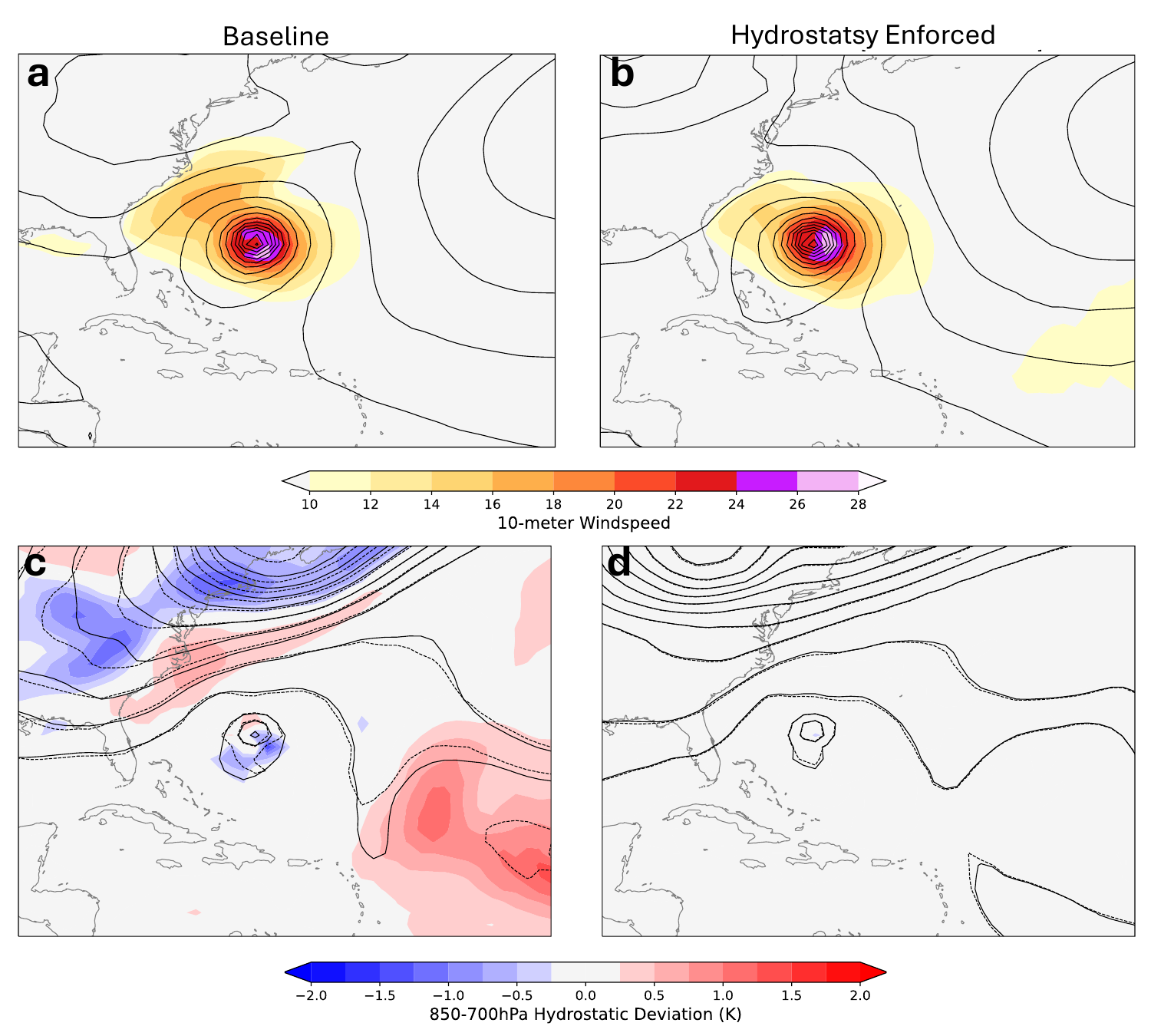}
  \centering
  \caption{Influence of the hydrostatic constraint on a 10-day forecast of hurricane Irma. 10-m wind speed (color) and 1000-hPa geopotential height (35-m contour interval) from the (a) baseline and (b) q50-constrained models. Hydrostatic deviation (color), together with simulated $T_v$ averaged across the 850-700-hPa layer (solid, 2 K intervals) and layer-average $T_v$ calculated from right side of (\ref{eqn:FD_hypsometric})  (dashed, 2 K interval) for the (c) baseline and (d) q50-constrained models.}
    \label{hydrostasy-irma}
\end{figure}

Figure~\ref{hydrostasy-irma}  compares 10-day baseline and q50 constrained forecasts, for hurricane Irma. These forecasts were initialized from ERA5 data on August 31, 2017, one day before Irma rapidly intensified into a hurricane. At the time shown in Fig.~\ref{hydrostasy-irma}, observations place Irma closer to the Florida keys, so both of these forecasts have her center too far to the northeast.  Ensemble forecasts would likely provide better guidance about the 10-day hurricane track, but that is not our current goal.  Rather, these forecasts are sufficiently similar to allow us to isolate differences arising from enforcing approximate hydrostatic balance.

The q50-constrained model generates stronger 10-m winds than the baseline on the eastern side of the coarsely resolved eyewall (data are plotted on a $1^{\circ}\times 1^{\circ}$ latitude-longitude mesh).  Also of note is the center of strong winds outside and northwest of the eye in the baseline forecast (Fig.~\ref{hydrostasy-irma}a). The strength and placement of these winds with respect to the eyewall is unusual. Indeed the weak horizontal gradient in the 1000-hPa height field  is too small for dynamical consistency with these strong 10-m wind speeds in our baseline forecast. The realism of the wind field is improved in the q50-constrained forecast (Fig.~\ref{hydrostasy-irma}b). Given the co-location of these unphysical winds with a large positive virtual temperature error (Fig.~\ref{hydrostasy-irma}c), this feature appears to have been ameliorated by the imposition of physical dynamical constraints.  Other regions of nontrivial imbalance are also present throughout the plotted domain in our baseline simulation.  In contrast, the q50-constrained forecast shows no regions of imbalance larger than 0.25 K (Fig.~\ref{hydrostasy-irma}d) and the $T_v$ contours from the forecast fields almost perfectly overlay those from the hydrostatic computation.

\bigskip

\section{Conclusions}
\label{sec:conclusions}

We added the physical constraint of hydrostatic balance within each vertical column. It was imposed as a soft constraint through finetuning a baseline model trained exclusively on MSE loss. Because of our coarse vertical resolution and the need to simultaneously minimize  MSE losses for the prognostic fields, training reduced the validation loss for the hydrostatic-constraint only in the lowest two layers.  Nevertheless, the model solution adjusted toward hydrostatic balance in all layers over the first 3--10 days of autoregressive rollout.  

The superiority of the physics-constrained model over the unconstrained baseline was evident by 7--10 days in most of the prognostic fields, even at levels where the hydrostatic validation loss did not appear to improve during training.  This suggests the absolute magnitude of a single-time-step error is less important than the error orthogonal to the hydrostatic manifold that develops during the forecast---highlighting the importance of physical constraints.

Hurricanes can be long-lived storms, and it was possible to compare physics-constrained and unconstrained forecasts for hurricane Irma at 10-day forecast lead time.  The unconstrained forecast was clearly out of balance in the 850-700 hPa layer and produced very strong winds well northwest of the eye, where they would not ordinarily be expected.  The physics constrained model showed good hydrostatic balance throughout the domain, avoided generating very strong winds in the region NW of the eye and produced slightly stronger extrema in the 10-m windspeed along the eastern eyewall.

The accuracy of our hydrostatic constraint can be improved using finer vertical resolution, and that is one avenue for future improvements.  Further investigation of the complex relation between validation loss during training and RMSE and imbalance errors during autoregressive forecasts is also merited.

\begin{ack}
This research was partially supported by the Office of Naval Research under grant N00014-24-12528. NCC was supported by a National Defense Science and Engineering Graduate Fellowship. WY was supported by the U.S. Department of Energy, Office of Science, Office of Advanced Scientific Computing Research, Department of Energy Computational Science Graduate Fellowship under Award Number DE-SC0025528.  Finally, this work benefited substantially from the barrier-free high quality ERA5 dataset provided by the ECMWF. This research used resources of the National Energy Research Scientific Computing Center (NERSC), a Department of Energy Office of Science User Facility using NERSC award BER-ERCAP-0023977.


\end{ack}


\bibliographystyle{plain}
\bibliography{hydro_references}
\medskip


\appendix

\section{Additional Data Information}
Table~\ref{full_variable_descriptions} Lists all the prognostic and prescribed variables used to train the DL\textit{ESy}M model including mean and standard deviation scaling values for each variable and the per variable Lagrange multipliers used in the loss function.

\begin{table}[htb]
    \centering
    \caption{{\bf Initialization fields representing Earth System for DL\textit{ESy}M.} The columns show the variable name, the abbreviated symbol for the variable, mean and standard deviation used to normalize the data in the dataloader and the per variable loss weight/Lagrange multiplier used to construct the full training loss.}
    \begin{tabular}{SlSlSlSlSl}
        \hline
        \textup{Variable Name}        & Symbol    & Mean    & Standard Deviation & Loss Weight \\
        \hline
        \rowcolor{gray!10}
        1000-hPa geopotential height & $Z_{1000}$        &   $937 m^2/s^2$  & $902  m^2/s^2$ & $0.0586$ \\
        850-hPa geopotential height  & $Z_{850}$         & $14250 m^2/s^2$  & $1207 m^2/s^2$ & $0.1478$ \\
        \rowcolor{gray!10}
        700-hPa geopotential height  & $Z_{700}$         & $29768 m^2/s^2$  & $1746 m^2/s^2$ & $0.2910$ \\
        500-hPa geopotential height  & $Z_{500}$         & $55511 m^2/s^2$  & $4221 m^2/s^2$ & $0.4553$ \\
        \rowcolor{gray!10}
        250-hPa geopotential height  & $Z_{250}$         & $103580 m^2/s^2$ & $4610 m^2/s^2$ & $0.8660$ \\
        100-hPa geopotential height  & $Z_{100}$         & $159739 m^2/s^2$ & $4143 m^2/s^2$ & $0.9225$ \\
        \rowcolor{gray!10}
        50-hPa geopotential height   & $Z_{50}$          & $200887 m^2/s^2$ & $3873 m^2/s^2$ & $0.8106$ \\
        2-m specific humidity        & $Q_{2m}$          & $1.375^{-2} kg/kg$ & $4.198^{-3} kg/kg$ & $0.0387$ \\
        \rowcolor{gray!10}
        850-hPa specific humidity    & $Q_{850}$         & $5.996^{-3} kg/kg$ & $4.198^{-3} kg/kg$ & $0.0113$ \\
        700-hPa specific humidity    & $Q_{700}$         & $3.178^{-3} kg/kg$ & $2.769^{-3} kg/kg$ & $0.0077$ \\
        \rowcolor{gray!10}
        500-hPa specific humidity    & $Q_{500}$         & $1.100^{-3} kg/kg$ & $1.218^{-3} kg/kg$ & $0.0055$ \\
        2-m temperature              & $T_{2m}$          & $287.4 K$        & $15.39 K$        & $0.1297$ \\
        \rowcolor{gray!10}
        850-hPa temperature          & $T_{850}$         & $281.1 K$        & $12.35 K$ &  $0.0730$ \\
        700-hPa temperature          & $T_{700}$         & $273.7 K$        & $11.52 K$ &  $0.0997$ \\
        \rowcolor{gray!10}
        500-hPa temperature          & $T_{500}$         & $258.5 K$        & $10.96 K$ &  $0.0973$ \\
        250-hPa temperature          & $T_{250}$         & $225.6 K$        & $7.31 K$ &  $0.0345$ \\
        \rowcolor{gray!10}
        100-hPa temperature          & $T_{100}$         & $204.8 K$        & $11.4 K$ & $0.0781$ \\
        50-hPa temperature           & $T_{50}$          & $211.3 K$        & $7.49 K$ & $0.0354$ \\
        \rowcolor{gray!10}
        total column water vapor     & $\mcvar{TCWV}$              & $24.1 kg/m$      & $16.7 kg/m$ & $0.0357$ \\
        10-m windspeed               & $\windspeed{10}$         & $6.14 m/s$       & $3.66 m/s$ & $0.0057$ \\
        \rowcolor{gray!10}
        850-hPa windspeed            & $\windspeed{850}$        & $8.38 m/s$       & $5.70 m/s$ & $0.0049$ \\
        700-hPa windspeed            & $\windspeed{700}$        & $9.51 m/s$       & $6.72 m/s$ & $0.0064$ \\
        \rowcolor{gray!10}
        500-hPa windspeed            & $\windspeed{500}$        & $12.9 m/s$       & $9.74 m/s$ & $0.0081$ \\
        250-hPa windspeed            & $\windspeed{250}$        & $22.0 m/s$       & $15.7 m/s$ & $0.0098$ \\
        \rowcolor{gray!10}
        100-hPa windspeed            & $\windspeed{100}$        & $15.6 m/s$    & $10.6 m/s$ & $0.0113$ \\
        50-hPa windspeed             & $\windspeed{50}$         & $12.0 m/s$    & $11.0 m/s$ & $0.0184$ \\
        \rowcolor{gray!10}
        land-sea fraction            & $F_{ls}$          & -    & - & - \\
        surface elevation            & $z$               & -    & - & - \\
        \hline
        \\
    \end{tabular}
    \label{full_variable_descriptions}
\end{table}

\section{Other experiments}
\label{sec:other-experiments}
We ran many other experiments with different hydrostatic finetuning parameters. Here, we show one of those to indicate the robustness of the hydrostatic finetuning to hyperparameter changes. Table~\ref{tab:alphas-v3-1} shows the hyperparameters used for these experiments. They differ from the ones reported in the main section mainly in the Lagrange multipliers for the hydrostatic loss terms. 

\begin{table}[htb]
    \centering
    \caption{Table outlining the different $\alpha_k$ values used for the three finetuning experiments with $p$ 0.5, 0.75 and 0.95. The last column also shows the Lagrange multiplier used for the hydrostatic loss for each vertical slab.}
    \begin{tabular}{SlSlSlSlSl}
        \hline
        \textup{Vertical Slab}        & $\left. \alpha_k \right| Q(0.5)$    & $\left. \alpha_k \right| Q(0.75)$    & $\left. \alpha_k \right| Q(0.95)$ & Loss Weight \\
        \hline
        \rowcolor{gray!10}
        850-700 hPa & $0.211$ & $0.475$ & $1.213$ & $10^{-4}$ \\
        700-500 hPa & $0.255$ & $0.510$ & $0.957$ & $10^{-4}$ \\
        \rowcolor{gray!10}
        500-250 hPa & $0.640$ & $1.216$ & $3.391$ & $10^{-4}$ \\
        250-100 hPa & $1.895$ & $2.757$ & $4.135$ & $10^{-4}$ \\
        \rowcolor{gray!10}
        100-50 hPa  & $0.676$ & $1.420$ & $2.773$ & $10^{-4}$ \\
        \hline
        \\
    \end{tabular}
    \label{tab:alphas-v3-1}
\end{table}

\begin{figure}[htb]
\includegraphics[width=0.32\textwidth, keepaspectratio]{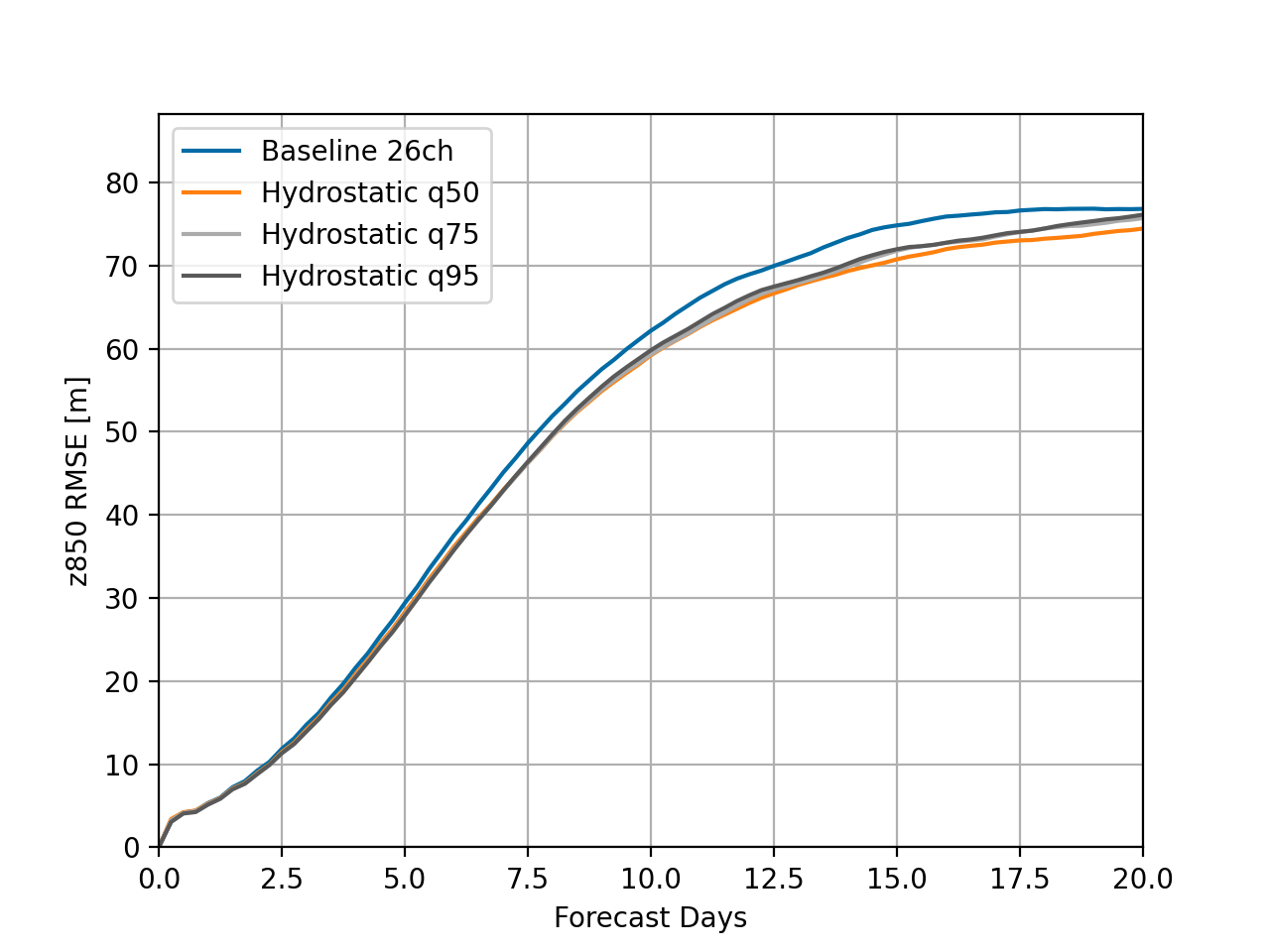}
\includegraphics[width=0.32\textwidth, keepaspectratio]{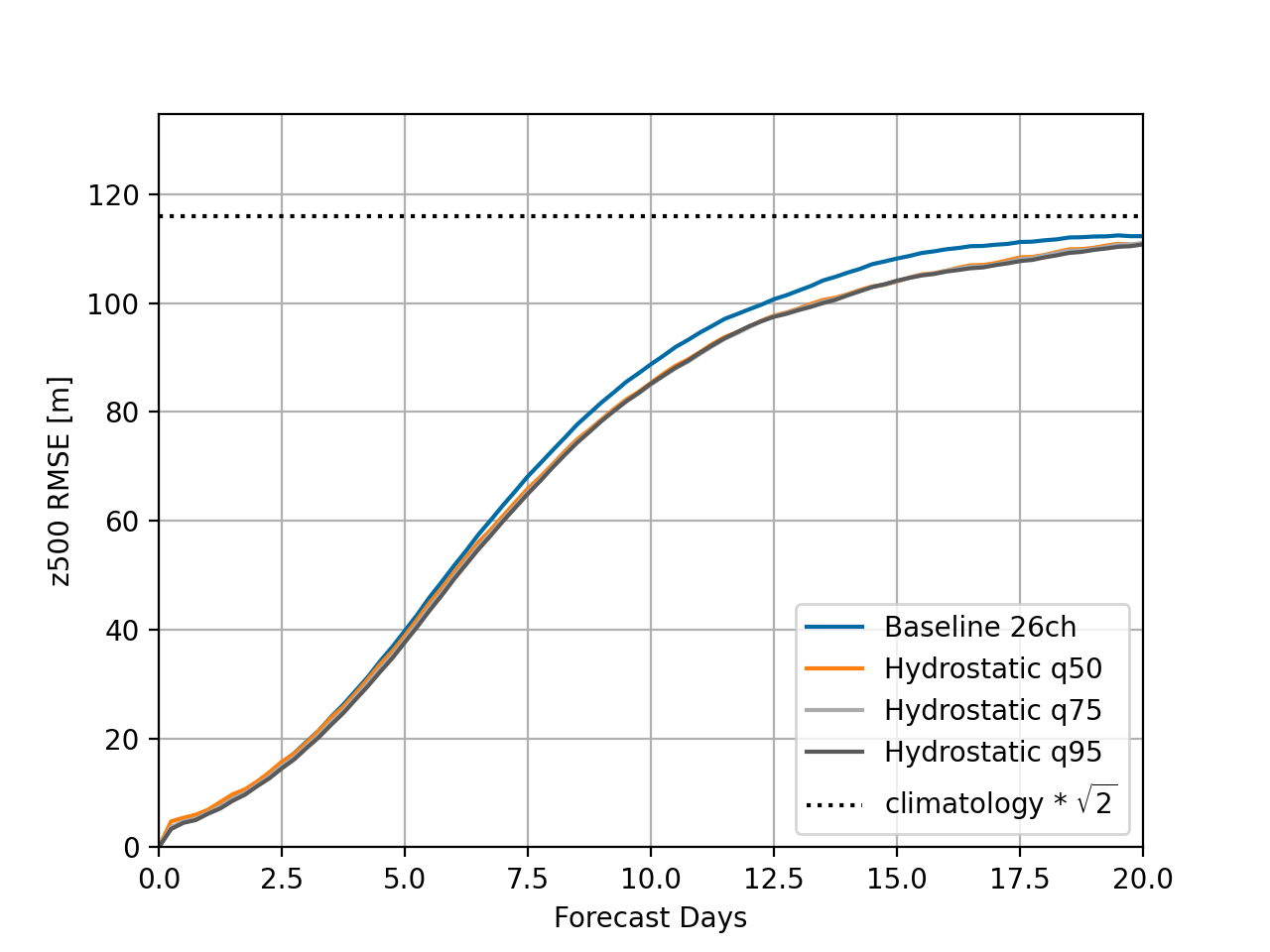}
\includegraphics[width=0.32\textwidth, keepaspectratio]{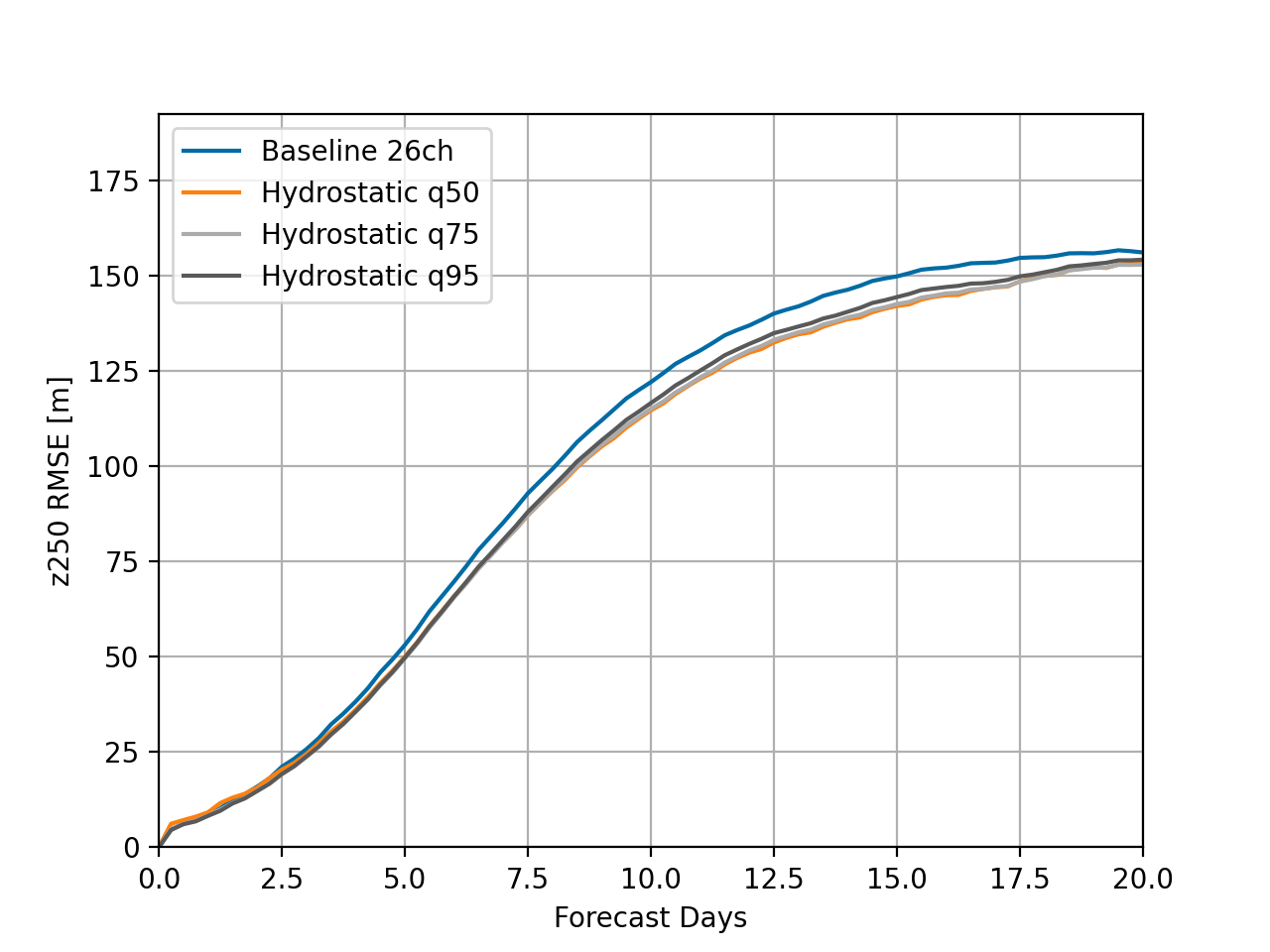}
  \centering
  \caption{Comparison of global RMSE curves between the baseline non-hydrostatic model and models with the hydrostatic constraint imposed above the 50, 75 and 95 percentiles. Variables are $Z_{850}$, $Z_{500}$, $Z_{250}$ (left to right).}
 \label{v3-1-rmse-curves}
\end{figure}

Figure~\ref{v3-1-rmse-curves} shows the geopotential RMSE curves evaluated for this version of the finetuned models similar to Section~\ref{sec:error-statistics}. The trends are very similar to results presented in Section~\ref{sec:error-statistics} with the hydrostatic constrained models outperforming the baseline model. This suggests that the results presented in Section~\ref{sec:error-statistics} are robust to checkpoint differences and hyperparameter changes and that the improvement in skill can be attributed to the hydrostatic finetuning.

\section{Validation of hydrostatic constraint implementation}

We validate our hydrostatic constraint implementation by considering a constant lapse rate temperature profile with a lapse rate of $-\mathrm{d}T / \mathrm{d}z = \Gamma$
\beq \frac{T}{T_0} = \left( \frac{p}{p_0} \right)^{\Gamma R / g} \eeq{eqn:constant-lapse-rate}
and setting $q = 0$. Given this analytical solution that is in hydrostatic balance, we can evaluate the accuracy of the numerical implementation.

\begin{figure}[htb]
\includegraphics[width=0.32\textwidth, keepaspectratio]{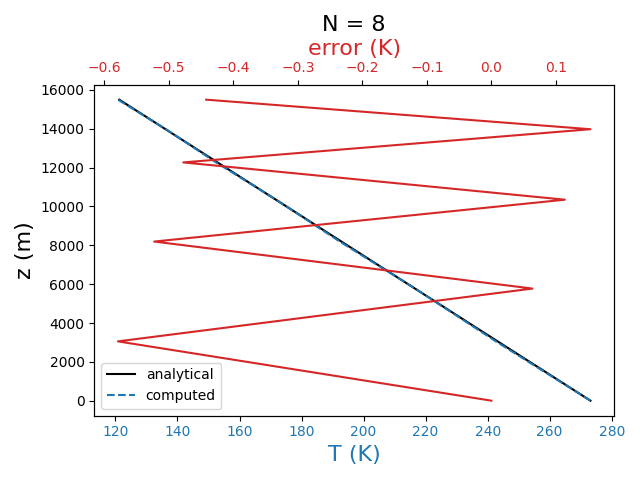}
\includegraphics[width=0.32\textwidth, keepaspectratio]{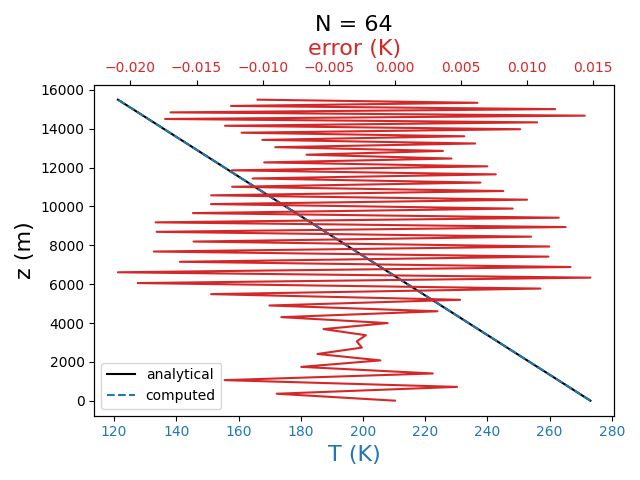}
\includegraphics[width=0.32\textwidth, keepaspectratio]{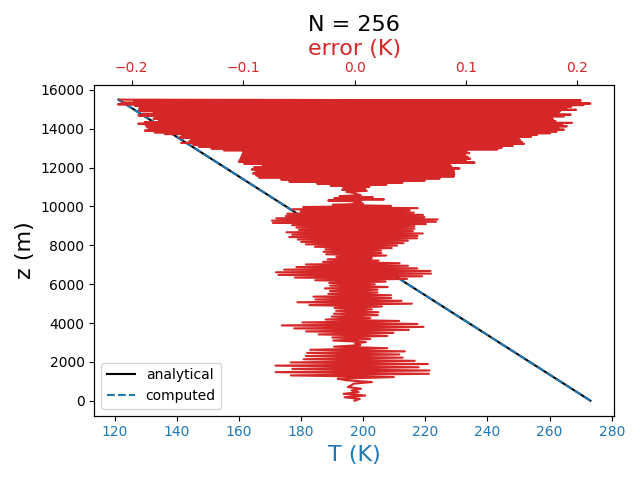}
\includegraphics[width=0.32\textwidth, keepaspectratio]{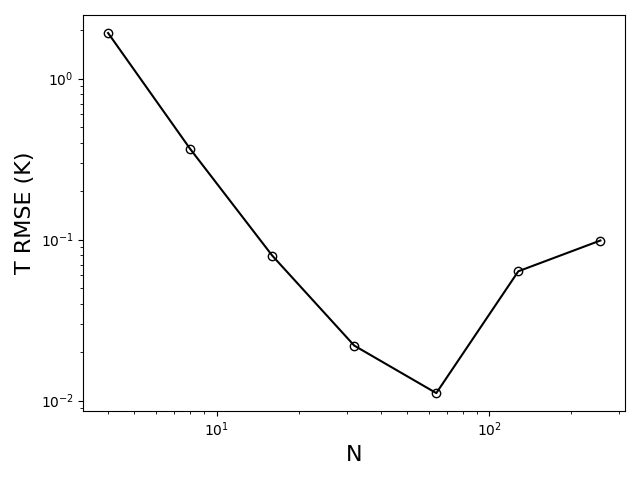}
  \centering
  \caption{$T_v$ error computed for a constant lapse rate analytical temperature profile for 8, 64 and 256 vertical levels (top, left to right). Bottom: $T_v$ error RMSE as a function of the number of vertical levels.}
 \label{Tv-validation-plots}
\end{figure}

Figure~\ref{Tv-validation-plots} shows $T_v$ errors computed for this analytical solution as a function of the vertical resolution. These errors reduce significantly with the number of vertical levels until 64 levels beyond which round-off error accumulation cause an increase in the error. This shows that the numerical implementation of the hydrostatic error computations are accurate within the limits of discretization and floating point errors.

\end{document}